# Quantum interference between independent solid-state single-photon sources separated by 300 km fiber

Xiang You[1,2,*], Ming-Yang Zheng[3,*], Si Chen[1,2,*], Run-Ze Liu[1,2,*], Jian Qin[1,2], M.-C. Xu[1,2], Z.-X. Ge[1,2], T.-H. Chung[1,2], Y.-K. Qiao[1,2], Y.-F. Jiang[3], H.-S. Zhong[1,2], M.-C. Chen[1,2], H. Wang[1,2], Y.-M. He[1,2], X.-P. Xie[3], H. Li[4], L.-X. You[4], C. Schneider[5,6], J. Yin[1,2], T.-Y. Chen[1,2], M. Benyoucef[7], Yong-Heng Huo[1,2], S. Höfling[5], Qiang Zhang[1,2,3], Chao-Yang Lu[1,2], Jian-Wei Pan[1,2]

[1] Hefei National Laboratory for Physical Sciences at Microscale and Department of Modern Physics, University of Science and Technology of China, Hefei, Anhui 230026, China.
[2] CAS Centre for Excellence in Quantum Information and Quantum Physics, University of Science and Technology of China, Shanghai, 201315, China.
[3] Jinan Institute of Quantum Technology, Jinan, Shandong, 250101, China.
[4] State Key Laboratory of Functional Materials for Informatics, Shanghai Institute of Microsystem and Information Technology (SIMIT), Chinese Academy of Sciences, 865 Changning Road, Shanghai 200050, China.
[5] Technische Physik, Physikalisches Instität and Wilhelm Conrad Röntgen-Center for Complex Material Systems, Universitat Würzburg, Am Hubland, D-97074 Würzburg, Germany.
[6] Institute of Physics, University of Oldenburg, 26129 Oldenburg, Germany.
[7] Institute of Nanostructure Technologies and Analytics, CINSaT, University of Kassel, Heinrich-Plett-Str. 40, 34132 Kassel, Germany.

---

**Abstract:**

In the quest to realize a scalable quantum network, semiconductor quantum dots (QDs) offer distinct advantages including high single-photon efficiency and indistinguishability, high repetition rate (tens of GHz with Purcell enhancement), interconnectivity with spin qubits, and a scalable on-chip platform. However, in the past two decades, the visibility of quantum interference between independent QDs rarely went beyond the classical limit of 50% and the distances were limited from a few meters to kilometers. Here, we report quantum interference between two single photons from independent QDs separated by 302 km optical fiber. The single photons are generated from resonantly driven single QDs deterministically coupled to microcavities. Quantum frequency conversions are used to eliminate the QD inhomogeneity and shift the emission wavelength to the telecommunication band. The observed interference visibility is 0.67±0.02 (0.93±0.04) without (with) temporal filtering. Feasible improvements can further extend the distance to ~600 km. Our work represents a key step to long-distance solid-state quantum networks.

**Introduction**

Quantum communications exploit the fundamental properties of quantum mechanics, such as superposition and entanglement, to implement communication tasks that are infeasible with the classical means. Examples include quantum key distribution[1,2] and quantum teleportation[3]. Since the early days of table top experiments[4,5], one of the most significant challenges of the field is to extend the distance of quantum communication to a practically useful scale. Exciting progress[6] has been made over the past decades that culminated at satellite-based quantum communication over thousand kilometers[7-12]. Taking advantage of the negligible photon loss in the empty outer space, the satellite-ground link has proven as an ultra-low-loss photonic channel.

In addition to the quantum channel, another important ingredient of the long-distance quantum communications is the quantum light source[13]. An ideal candidate is a single-photon source that emits one and only one photon each time[14-16]. To obtain a high count rate after transmission, the single-photon sources should have a high system efficiency (which includes the generation[17], extraction[18,19], and collection[20] efficiencies) and high repetition rate[21,22] (which is intrinsically limited by the emitter's radiative lifetime). For quantum network applications, such as quantum teleportation that requires interfering independent photons, the single photons should be transform limited[23]. Additional requirements include a scalable platform, tunable and narrowband linewidth (favorable for temporal synchronization) and interconnectivity with matter qubits. Quantum dots (QDs) have been considered a promising solid-state system for quantum networks[14,23,24]. However, previous attempts on quantum key distribution[25-27] with QDs were up to a few kilometers only, and QD-based two-photon interferences[28-35] were also limited to kilometer scale and mostly below 50% visibility. There are a number of challenges to achieve a long-distance quantum interference, including high performances on single-photon source brightness, purity, indistinguishability, wavelength band and matching, high-fidelity transmission, and more crucially, integrating all these parameters together compatibly.

In this Article, we report high-visibility quantum interference between two independent QDs linked with ~300 km optical fibers by developing efficient and indistinguishable single-photon sources, ultra-low-noise and tunable single-photon frequency conversion, and low-dispersion long fiber transmission. As a first step, our experiment points to a promising route to long-distance solid-state quantum networks.

**Single-photon sources**

Our experimental configuration is shown in Fig. 1. Two QDs are housed inside two cryogenic-free cryostats with a temperature of 4 K and 1.7 K, respectively. To maximize the efficiency and indistinguishability of the single photons, the QDs are spectrally and spatially optimally coupled to microcavities. As in our previous work[20], two different types of microcavities are used: QD1 is embedded inside a narrowband micropillar[36], and QD2 is coupled to a broadband bullseye cavity[20,37,38]. Under resonant π-pulse excitation by an ultrafast laser, resonance fluorescence single photons at wavelengths of $\lambda_{QD1}$ =893.16 nm ($\lambda_{QD2}$ =891.92 nm) are emitted from QD1 (QD2) and collected into single-mode fibers.

Under an 80.3-MHz pumping rate, at the output of collection single-mode optical fibers, the final single-photon rate is 20.2 MHz and 16.2 MHz for QD1 and QD2, respectively, corresponding to a system efficiency of 25% and 20%. The second-order correlations of the single-photon sources are characterized by Hanbury-Brown-Twiss measurements which give $g^2_{QD1}(0) = 0.072(1)$ and $g^2_{QD2}(0) = 0.051(1)$, as plotted in Fig. 2a. The mutual indistinguishability between two single photons from the same QDs is measured using a Hong-Ou-Mandel interferometer where they overlap at a 50:50 beam splitter.

These two single photons are consecutively emitted from the same QDs with a time separation of 12.5 ns. Figure 2b shows the histograms of normalized coincidences for the two photons set at parallel and orthogonal polarizations. After correction of the residual second-order correlation, we extract a photon indistinguishability of 91.9(1)% and 83.9(3)% for QD1 and QD2, respectively.

It is important to note the difference between the mutual indistinguishability at 12.5-ns separation and Fourier transform limit[23]. The former is immune to any environmentally induced spectral diffusion that occurs at a time scale much slower than 12.5 ns. What really matters for the quantum interference between independent QDs is the degree of transform limit, that is, the ratio of $T_2/2T_1$, where $T_1$ and $T_2$ are radiative lifetime and coherent time of the single photons, respectively. We measure $T_1$ using time-resolved pulsed resonance fluorescence. By fitting the exponential decay, we extract the radiative lifetime $T_1$ of 78.0(1) ps for QD1 and 69.9(1) ps for QD2, as illustrated in the insets of Fig. 2c. The coherence time is measured using a Mach-Zehnder interferometer and then calculated by fitting the fringe contrast as a function of temporal delay. Fig. 2c shows the coherence time of the single photons, which is 126(1) ps for QD1 and 105(2) ps for QD2. These allows us to calculate the degree of transform limit as 80.8(1)% for QD1 and 75.1(1)% for QD2, which are slightly lower than the 12.5-ns indistinguishability as we expected.

**Quantum frequency conversion**

There are two major challenges in sending the QD single photons through long-distance optical fibers and observing quantum interference. First, the InAs QDs emission is at a wavelength of ~890 nm, which should be converted to telecommunication wavelength to exploit the low transmission loss in commercially available fibers. So far, the QDs directly emitting single photons in the telecommunications wavelength[39-43] have not yet reached a performance comparable to their near-infrared counterparts. Second, the self-assembled QDs emit single photons intrinsically at different wavelength, which would reveal which-way information to prevent the Hong-Ou-Mandel interference.

In this work, we use quantum frequency conversion[44-46] to overcome both problems. To this end, we fabricate Periodically Poled Lithium Niobate (PPLN) waveguide for difference frequency generation (see Fig. S3). The energy conservation demands $1/\lambda_c=1/\lambda_s-1/\lambda_p$, where $\lambda_s$, $\lambda_p$ and $\lambda_c$ represent the wavelengths of the signal, pump, and converted photons, respectively. To precisely tune the two converted wavelengths into resonance, the pump lasers have both a coarse tuning range of ~1 nm and a fine tuning resolution of 3.6 MHz using the laser PZT actuator, which is ~40 times and ~0.1% of the QD emission linewidth, respectively (Fig. 3a). For the wavelengths of QD1 and QD2, the pump lasers are tuned at 2049.98 nm and 2043.46 nm, respectively, which convert both into 1582.75 nm (as labelled in Fig. 1).

By optimizing the nonlinear interaction, waveguide coupling, and transmission rate, the overall single-photon conversion efficiencies reach ~50% for both devices (Fig. 3b). To suppress the noise background from the residual pump laser, harmonic generation, and broadband Raman photons induced by the strong pump laser, we use a combination of dichromic mirrors and optical filters to obtain a signal-to-noise ratio of 28-30 dB (Fig, 3c). We note that an advantage of the frequency conversion process is that it does not interfere with the quantum emitter itself. To test whether the converted photons still preserve the coherence properties of the signal single photons, we measure the purity and coherence time of the single photons after conversion, which, as plotted in Fig. S4 and Fig. 2c, show near-perfect overlap with the data before conversion.

### **Fiber transmission of single photons**

The dominant loss is from the long-distance fiber transmission of the single photons. As the transmission rate of the fiber is 0.19 dB/km, the loss over 300 km is 57 dB. Fiber transmission of single photons not only causes photon loss, but can also influence photon's properties. For example, the orientation of photon polarization can be changed in optical fibers. The photon's arrival time can drift due to the change of the fiber length caused by temperature fluctuation.

Efforts are taken to preserve the photon's properties during the fiber transmission. To reduce the drift of photon's arrival time, the temperature of the fibers is stabilized within $\pm0.1$ degree. The measured typical time drift is within 10 ps per hour, which is much smaller than the photon's coherence time. A set of half- and quarter-wave-plates is used to control the polarization. As shown in Fig. S6, there is a slow wandering of polarization drift over hours, which is transformed into ~10% level efficiency loss by applying polarization filtering at the end of the optical fibers.

There is also an effect of frequency dispersion in optical fibers owing to a wavelength-dependent velocity, which could reduce the indistinguishability of the single photons. The dispersions of QD1 and QD2 single photons over the 150-km fiber are 66.5 ps and 89.4 ps, respectively, which are comparable to the single photon's coherence time of 105-120 ps. However, if we arrange the identical single photons to go through the same fiber length, the photons will experience the same dispersion and thus remains identical. Therefore, the symmetric transmission configuration set-up in our experiment makes the two-photon interference immune to fiber dispersion[47,48].

**Remote two-photon interference**

After faithful transmission over the optical fibers, the two single photons in the outputs are synchronized and superposed on a beam splitter for quantum interference. We use superconducting nanowire single-photon detectors with an efficiency of 76% and a time resolution of ~70 ps to register the finally arrived photons. The two-photon coincidence counts when the two photons are controllably set at resonant (red) and far-tuned (black, $\Delta\nu$=38 GHz) are plotted for a range of total fiber length of 24 m (Fig. 4a), 101 km (Fig. 4b), 201 km (Fig. 4c) and 302 km (Fig. 4d). Note that the counts presented here are the raw data without any background subtraction.

The extracted raw visibilities are at a level of 0.67±0.02 for different optical fiber length. There is no evident drop in the visibility for increasing fiber length, as expected from the dispersion-cancellation symmetric transmission. These raw visibilities significantly exceed the classical limit of 50% which conclusively demonstrate genuine two-photon quantum interference. The visibilities are plotted (red dots) in Fig. 4e as a function of the fiber length, which is in good agreement with the theoretical calculation (red line) that considers the $T_2/2T_1$ for each QD, their bandwidth mismatch, and their imperfect second-order correlations. Considering the $g^2_{QD1}(0)$ and $g^2_{QD2}(0)$, the corrected two-photon interference visibility is 0.73±0.02 at 302 km.

Temporal filtering can also significantly increase the two-photon interference visibility. The time resolution of the single-photon detectors is 70 ps, much smaller than photon's coherence time. We plot in Fig. 4f the raw visibilities as a function of coincidence time window. The raw visibility increases substantially with narrowing time window, as the temporal filtering effectively improves the coherence of the single photons. At 20 ps, the visibility reaches 0.93±0.04. Note that such a filtering, only at the cost of heralding efficiency, can be useful in future experiments on high-fidelity entanglement swapping of single photons[49,50] and single spins[51,52].

**The future**

Figure 5a summarizes two-photon interference distance and visibilities of previously reported work between two QDs, to the best of our knowledge[28-35,53]. This experiment establishes a distance that is more than 2 orders of magnitude larger than the previous record, with simultaneously the highest visibility.

A number of straightforward improvements can further extend the distances. The short $T_1$ of the QDs enabled by the high Purcell factors allows to increase the repetition rate from 80 MHz to 2.5 GHz, a ~30 times enhancement. Using tunable open microcavities[22], the single photon system efficiency is feasible to reach 80%. In addition, ultra-low-loss optical fiber with transmission loss of 0.16 dB/km has become available. A numerical simulation curve is plotted in Fig. 5b. With these readily improvement, the transmission distance can be extended to ~600 km where the coincidence count rate will be 0.012 Hz with a signal-to-noise ratio of 10 dB. Such a distance scale is already comparable to the well-developed twin-field quantum key distribution experiments[54,55].

In summary, our work represents an important step toward quantum telecommunication networks using semiconductor QDs and telecom fiber channels. The experiment creates a solid-state platform to implement quantum teleportation[5], entanglement swapping[49,50], quantum relay[56], and teleportation of controlled-NOT gates at hundreds of kilometers scale in a multi-user network configuration. A key advantage of using the single QDs, compared to spontaneous parametric down-conversion, is the intrinsically deterministic single-photon emission and natural suppression of double pair events, which can allow the realization of multi-photon entanglement and interferometry in a non-post-selection way[6]. A large number of entangled photons can be generated in this platform by, for example, heralded creation of three-photon Greenberger-Horne-Zeilinger states[57] from six single photons, and using it as a basic resource and fuse into larger ones[58], which will be useful resources for all-photonic quantum repeaters[59] and distributed quantum computing. The distances and functionalities can be further improved by combining with suitable quantum memories. Thus, inter-city-scale fully quantum networks appears technologically promising based on a scalable semiconductor platform.

**Figure Caption**

**Fig. 1 | Experimental configuration of quantum interference between two independent solid-state QD single-photon sources separated by 302 km fiber.** Both QDs are embedded in microcavities, with QD1 in a micropillar and QD2 in a bullseye cavity. Under resonant π-pulse excitation (not shown), the single photons are emitted from QD1(QD2), collected by a confocal setup, and then sent into QFC1(QFC2) which consists of PPLN-WG, pump laser (not shown) and filters (DM: dichromatic mirror, LP: Long Pass, BP: Band Pass). The wavelength of single photons from QD1 (QD2) is converted from near-infrared to 1582.75 nm in QFC1 (QFC2) by adjusting wavelengths of pump lasers. The down-converted photons both transmit through 151 km optical fiber and impinge upon a 50:50 beam-splitter (BS) via Hong-Ou-Mandel interference. Arrivals of single photons after interference are detected by two superconducting nanowire single-photon detectors (SNSPDs) and then analyzed using a time-to-digital converter (not shown). The emissions of single photons from QD1 and QD2 are temporally synchronized by pumping with the same laser. The combination of a half-wave-plate (HWP), a quarter-wave-plate (QWP) and a polarization beam splitter (PBS) makes sure the two single photons will have the same polarization during interference. All fibers are single mode to transform photons into the fundamental transverse Gaussian mode for good spatial-mode overlap.

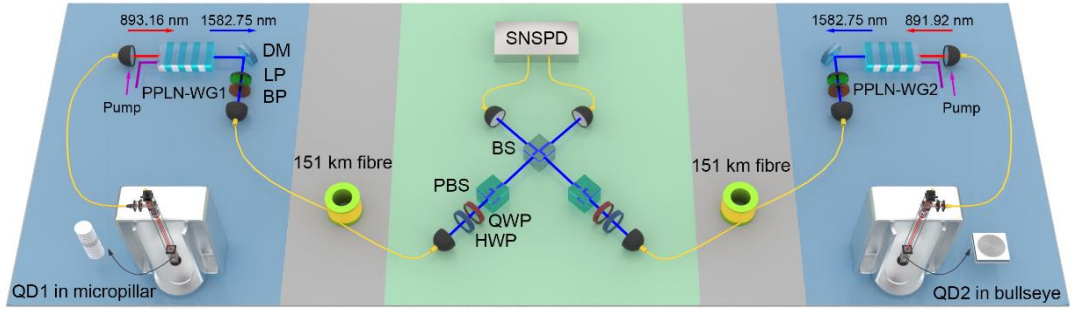

**Fig. 2 | Characterization of single photons emitted from QD1 and QD2, respectively. a**, Single photon purity: Hanbury-Brown-Twiss (HBT) measurements give $g^2_{QD1}(0) = 0.072(1)$ and $g^2_{QD2}(0) = 0.051(1)$. **b**, Indistinguishability: Hong-Ou-Mandel (HOM) measurements give calculated indistinguishability of 91.9(1)% for QD1 and 83.9(3)% for QD2 after correction. The red (black) data are normalized coincidence counts for polarization parallel (orthogonal) two photons. **c**, Coherence time: Measurements are carried out using Mach-Zehnder interferometer both before QFC1 (QFC2) and after QFC1 (QFC2). By fitting the fringe contrast as temporal delay, we get extracted coherence time of 126(1) (105(2)) ps and 123(3) (103(2)) ps at different positions for QD1 (QD2). The insets show the corresponding single photon radiative lifetimes for QD1 and QD2, which are calculated by fitting the one-sided exponential decay.

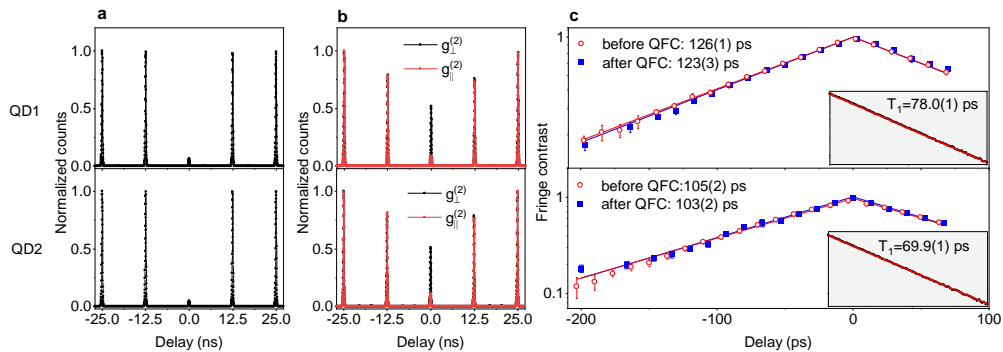

**Fig. 3 | Characterization of the quantum frequency conversion (QFC) setup. a**, Fine tuning of the wavelength of down-converted photons as a function of the position of pump laser's PZT actuator. The tuning resolution is ~0.03 pm, corresponding to ~3.6 MHz in frequency domain. The conversion efficiency is stable in the whole fine tuning range. **b**, Conversion efficiency and **c**, signal-to-noise ratio as a function of pump power. The maximum end-to-end efficiency is 48% (52%) at 271 mW (461 mW) for QFC1 (QFC2). The corresponding signal-to-noise values at

maximum efficiencies are 29.8 dB and 28.5 dB, respectively.

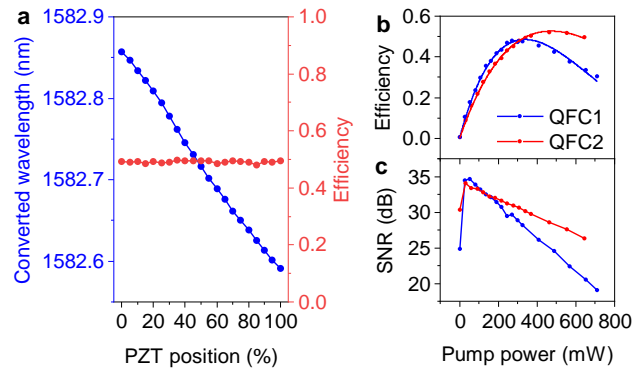

**Fig. 4 | Quantum interference between two solid-state QD single photon sources. a-d**, Measurements of coincidence counts between two down-converted photons separated by total fiber length of 24 m, 101 km, 201 km and 302 km, respectively (The 24 m is from the photon collection system before QFC). Red and black dots are the two-photon coincidence counts under spectrally resonant and 38-GHz detuned excitations, respectively. **e,** Experimental raw visibilities and theoretical visibility (red line) as a function of fiber length. Both are well above the classical limit of 50%. **f,** Dependence of raw visibility on coincidence time window with experimental data extracted from **d**. The raw visibility reaches up to 0.93±0.04 at 20 ps.

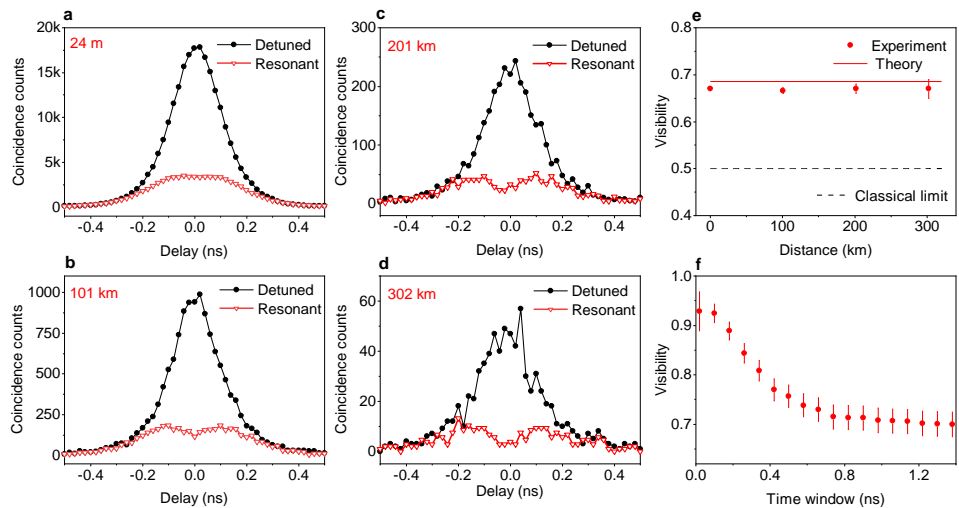

**Fig. 5 | Summary of previously reported work and outlook. a,** Summary of quantum interference visibilities between two solid-state QD single-photon sources as a function of distance. **b,** Simulations of two-photon coincidence count rate and signal-to-noise ratio as a function of optical fiber length with different system

parameters. The solid lines are simulated with parameters of this experiment, including $\upsilon_{pulse}$=80 MHz (repetition rate of pulsed excitation laser), $\eta_{sys}$=0.2 (photon system efficiency), $n_{dc}$=300 Hz (dark counts of SNSPD) and $\eta_{loss}$=0.19 dB/km (loss of optical fiber). The dotted lines are simulated with feasibly improved parameters of $\upsilon_{pulse}$=2.6 GHz, $\eta_{sys}$=0.8 and $\eta_{loss}$=0.16 dB/km.

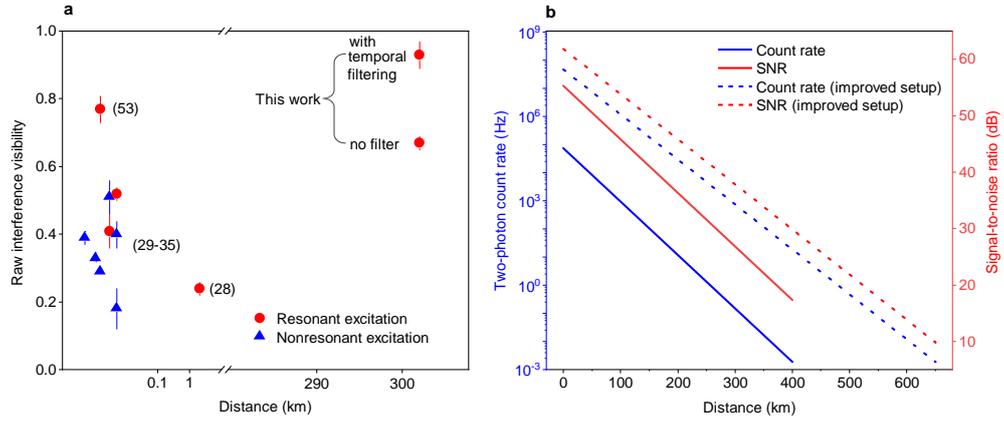